# A Wireless Sensor Network based approach to monitor and control air Pollution in large urban areas

Seyed Pourya Miralavy[1], Reza Ebrahimi Atani[2], Navid Khoshrouz[2]

[1]Department of Computer Science and Engineering, Shahid Beheshti University, Tehran, Iran.
[2]Department of Computer Engineering, University of Guilan, Rasht, Iran.
i.miralavy@sbu.ac.ir, rebrahimi@guilan.ac.ir, navid.khoshrouz@gmail.com

*Abstract*: Air pollution is a major concern in large urban areas. Studies show that about 78 percent of the air pollution in large cities are due to the exhaust emission of on-road or non-road vehicles. Various studies have been made to monitor and control the pollution level, emitted by the vehicles but some main factors like ease of implementation or feasibility of the proposed approaches are not focused in the told studies. In this paper we propose a wireless sensor network solution to monitor vehicles' exhaust system's pollution emission. Furthermore, we evaluate the feasibility of our proposed approach with respect to energy consumption and network lifetime by means of simulation. Our results indicate that our proposed scheme is practical for implementation using well known MAC layer protocols.

## 1. Introduction

Wireless Sensor Networks (WSNs) are becoming increasingly attractive for a variety of application areas, including industrial automation, security, weather analysis, and a broad range of military scenarios. Wireless sensor networks are dense wireless networks of sensor nodes collecting and disseminating environmental data. Sensor nodes are small low-power devices constrained severely in their computation, communication, and storage capabilities, usually for economical reasons. They may sense around themselves, communicate over wireless channels within short ranges, and frequently fall into the sleep mode for saving their power. A sensor node typically contains a power unit, a sensing unit, a processing unit, a storage unit, and a wireless transmitter / receiver. Figure 1 shows the hardware architecture of a sensor node [21-28].

Nowadays automobiles play an integral role in comforting people's life by providing ease of mobility. However, they cause air pollution when the level of exhaust gas emission from vehicles, breaches the standard levels [15] due to the incomplete combustion of fuel, supplied to engines [18]. Figure 1 shows that 78% of the air pollution comes from on-road and non-road vehicles. Studies conducted from 1982 to 1998 indicate that, even short-term exposures to sulfur-oxide in polluted air, might result in cardiopulmonary diseases [4]. This shows the substantiality of the air pollution problem; and thus has resulted in numerous researches to monitor and control air pollution, especially in large urban areas. In recent years, using interdisciplinary approaches has become more widespread to find efficient solutions for environmental problems. Developments in the field of computer science and especially wireless sensor networks has resulted in introduction of novel intelligent frameworks which are to be used to gather environmental data [11].

The rest of this paper is organized as following: Next section describes the significance of the air pollution monitoring and the goal of this research. After that we report a review of the related works. Then we describe each MAC layer protocols, used in our study for pollution monitoring systems as well as introducing our proposed method and hardware. Next, we discuss the simulation platform set to evaluate the performance of the described pollution monitoring approach and we explain our derived results. Finally, in the last section we conclude our work and talk about the future research possibilities in this matter.

## 2. SIGNIFICANCE OF AIR POLLUTION MONITORING

Rapid growth in Asian countries due to urbanization, motorization, and economic growth has raised serious concerns about the air pollution problem. Increase in mortality and respiratory diseases are the main outcome of the addition in air pollution in these countries. Figure 2 shows the relationship between Air Quality Index (AQI) [2014] and 53 literature results on mortality system and diseases of circulatory system: pneumonia, COPD, cardiovascular, cerebrovascular, ischemic stroke, respiratory disease, CE, CI, MI, ICB, IHD) in Asian cities (Hong Kong, Seoul, Inchon, Taipei, Bangkok, Gaoshiung, Shanghai, Beijing, Delhi, 13 cities in Japan, Tokyo) from 1999 to 2006. Annual value of PM10, TSP, SPM, $NO_2$ and $SO_2$ of these cities are converted to AQI according to U.S. EPA regulations. The linear indicates direct relationship between AQI and the probability of mortality. Furthermore, Figure 3 shows the relationship between AQI and hospital admissions of these countries. Similar to Figure 2, higher AQI results in higher hospital admissions [9].

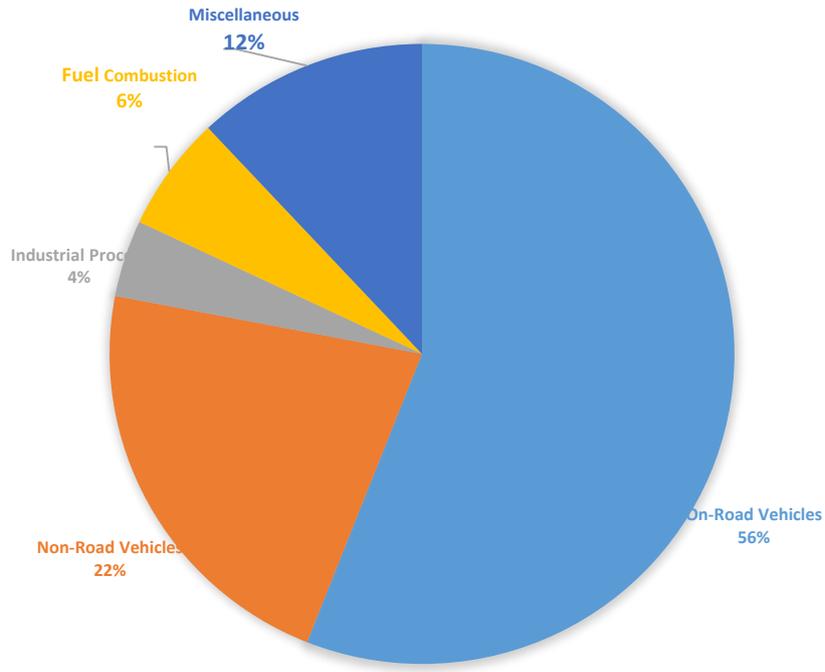

**Figure 1:** Sources of Air Pollution [3].

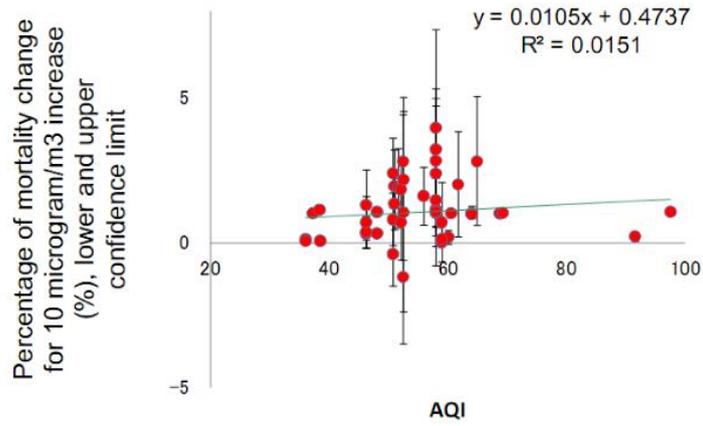

**Figure 2:** AQI and Mortality [9].

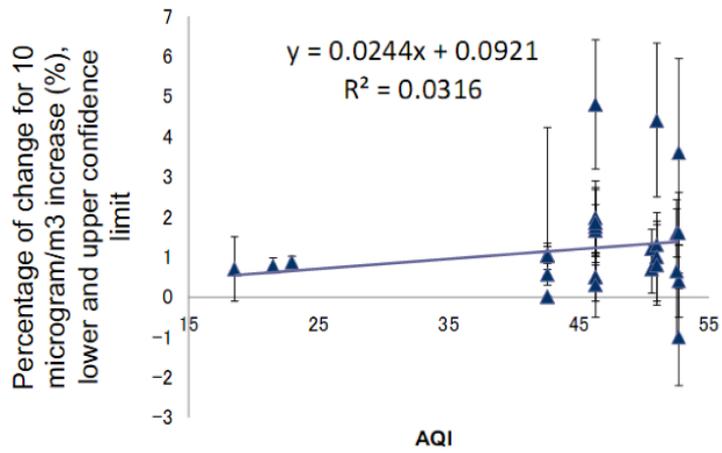

**Figure 3:** AQI and Hospital Admissions [9].

As it is discussed in the previous section, a considerable portion of this pollution is caused by on-road and non-road vehicles. In large cities like Tehran, Iran, schemes like odd-even rationing [17] has been applied to reduce the negative pollution effect of these sources [6].

## 3. RELATED WORKS

Nowadays with the development of technology, many environmental scientists use the means of computer science to solve available challenges. These challenges include a broad range of environmental problems.
For instance, Xu et al. explored the abilities of wireless sensor networks to monitor marine environments instead of traditional expensive oceanographic vessels [8]. In 2013, Lazarescu et al. suggested a WSN platform for long-term environmental monitoring purposes which takes the approach of Internet of Things into account [16].

Monitoring vehicular emission has also been an active research interest in recent years. Some researchers suggested using gas sensors in vehicles' exhaust system. They proposed using a GPS module in the node to show the vehicle-owner, the nearest service stations [3]. The goal of those researches was more focused on informing the vehicle-owner about the engine issue than to monitor the whole impact of vehicle exhaust emission on the pollution level of the city or environment.

Evaluating the performance of MAC layer protocols in different wireless sensor network environments such as underwater environment [5], or with a focus on different metrics such as power consumption [12] has also been an interesting research topic for many researchers. Considering the nature of WSNs, reducing the power consumption in such networks has extensively been under research and study [19].

The main challenge of Wireless Sensor Networks is the network's energy consumption and in other words, the lifetime of the network. Although related conducted researches have already focused on this issue, it is yet possible to further improve this significant challenge to have environmental monitoring systems with longer lifetimes.

## 4. PROPOSED SCHEME AND STUDY PROTOCOLS

In this section we first introduce our proposed scheme which is inspired from the work proposed in [3] and give a summary of the study MAC Layer Protocols.

Wireless Sensor Networks (WSN) are categorized as a type of computer networks which are typically consisted of several to thousands of low-cost, independent sensor nodes, which are used to sense, process and transmit environmental data. They send their gathered data, using wireless technology to one or several base stations to further process the received data and use it towards reaching the system goals and applications [14].

A possible approach to control the pollution problem is to monitor the exhaust emission of vehicles in large cities. Identifying vehicles which emit pollutant gases more than a standard level, can be used to alarm the owners to have their vehicle serviced; this will reduce the gas emission which in turn helps decreasing owners' gas expenses. If the problem was to continue, having such systems enables authorities to take proper actions to control the air pollution. As the air pollution problem intensifies, rules for restricting further pollution and reducing it, becomes stricter. For instance, some rules have already been applied in several countries like Japan [1] or United States. According to United States Environmental Protection Agency (EPA), "The 1990 Act also established the requirement that passenger vehicles be equipped with on board diagnostics" [2].

Although, it is possible to implement a built-in sensory system to measure vehicles' exhaust emission to modern vehicles, it is a quite time-consuming, expensive and rigorous process for older vehicles which are actually the main source of air pollution. To monitor and control vehicles' exhaust emission, in order to solve the environmental issue of air pollution, authorities have to monitor all the vehicles. Therefore, in this paper, we intend to study the feasibility of implementing a totally independent sensor array node on vehicles' exhaust system which uses an AA battery as its source of energy by simulating a city, in which vehicles take random routes based on Manhattan mobility model; and transmit sensory data to a base station located in the middle of the city. Fig. 2 gives a gist of the simulation area and our intended work. We then compare the scenario results for different widely used MAC layer protocols to discover the best fit for our proposed approach.

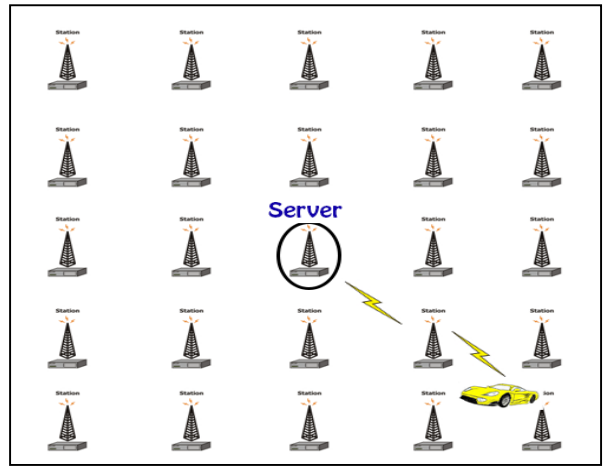

**Figure 4:** Wireless mesh scenario used for the simulation

**MAC LAYER PROTOCOLS:**

**SMAC:** SMAC [10] is a MAC protocol based on sleep-listen schedules. Two neighboring nodes wake up at the same time to listen or transmit their data. One of the main problems of SMAC is unsynchronized schedules of neighboring nodes, which result in more energy consumption while having idle listening or overhearing nodes.

**TDMA:** A TDMA-based (Time Division Multiple Access) MAC protocol [13] allocates different time slots for nodes to send or receive packets. These time slots are called a TDMA frame. In this simulation, we have used a preamble-based TDMA MAC protocol in which each TDMA frame contains preambles beside the data transmission slot.

**IEEE 802.11:** IEEE 802.11 MAC protocol is based on Carrier Sense Multiple Access / Collision Avoidance (CSMA/CA) technique for wireless LANs.

**IEEE 802.15.4:** IEEE 802.15.4 MAC protocol [11] is introduced as a low-power type of 802.11 for wireless LANs. It is widely used for specification of low-data-rate wireless transceiver technology, which has long battery life, and low complexity.

## 5. PROPOSED SCHEME:

In our proposed scheme we suggest implementing a sensor node on vehicles' exhaust system to monitor the emitted gases. This sensor node should be independent from the vehicles' built-in Environmental Control Unit (ECU) to make it possible for the vehicle owners to implement it with ease and in not much time and without modifying their vehicles' electronics.

Sensed data from the sensor nodes in addition to vehicles' information are then sent to a base station located in the large urban areas and controlled by the authorities. In order to do so, availability of wireless infrastructures to receive and forward the data is a necessity.

If the level of the pollution in the sensed data exceeds a certain standard level, authorities inform the vehicle owner to solve the engine issue of the vehicle and if the problem was to continue, charges will be set for the vehicle owner.

## 6. SIMULATION SCENARIO AND RESULTS

To evaluate the energy efficiency and other performance factors of the MAC layer protocols used for the system under study we set up a simulation. As it is shown in Figure 4, while a wireless mesh topology is assumed for the base stations' communication, vehicle nodes follow Manhattan mobility model in our simulation.

The vehicle sensor node sends data gathered from its exhaust system to a nearby station using AODV routing algorithm. This data transmission occurs in a timely interval which in this simulation is defined much lower than the real values in order to have meaningful graphs to analyze.

Table1. NS2 Simulation Parameters

| Parameter | Value |
|---|---|
| Channel Type | Wireless Channel |
| Propagation Model | Two Ray Ground |
| Network Interface | Wireless Phy |
| MAC type | 802.11/SMAC/TDMA/802.15.4 |
| Number of Base stations | 25 |
| Number of vehicles | 1 |
| Routing Protocol | AODV |
| Topology Size | 1000x1000 meters |
| TX Power | 2W |
| Node Initial Energy | 4700 J |
| Antenna Type | Omni directional |
| IFQ | DropTail/PriQueue |
| IFQ Length | 50 bits |
| Traffic Generator | CBR |
| CBR Interval | 0.1 sec |
| CBR Packet Size | 512 bytes |
| Mobility pattern | Manhattan Model |
| Simulation Time | 600 sec |
| Number of Simulations | 10 |

To perform the simulation NS2 simulator [20] are performed in a Linux Ubuntu 14.04 64-bit system (processor of Intel® CoreTM i7-5500U CPU @ 2.4 GHz 2.4 GHz, RAM of 8 GB DDRIII and GPU of NVIDIA GeForce 860M 2GB). Table I. shows the parameters set for the simulation scenario.

We have investigated the performance of the system based on the energy consumption, delay and packet delivery rate metrics.

**Residual Energy:** We aim to analyze vehicle node's remaining energy at the end of the simulation using different MAC layer protocols. In order to do that as mentioned in section IV a Manhattan vehicle mobility pattern is used.

The average simulation results taken from 10 runs, using IEEE 802.15.4, IEEE 802.11, SMAC and TDMA MAC protocols are presented in Figure 4. As it is shown, TDMA outperforms the rest of the MAC protocols with respect to the remaining energy. As it can be observed in Table 2 TDMA and SMAC MAC layer protocols have sent lower number of packets comparing to 802.11 and 802.15.4. This saves a lot energy which is used for transmission purposes. While it might justify TDMA MAC protocol's good performance here, it does not justify SMAC's bad performance. Unsynchronized schedules of neighboring nodes in SMAC protocol leads to wasting energy which is the main reason of this protocol's bad performance in this metric.

**Delay:** Delay is measured as the packet travel time between the moving nodes to the server. Average delay results are shown in Figure 5. It can be seen that while SMAC has the worst performance with average delay of 2.81ms, IEEE 802.11 and IEEE 802.15.4 outperforms other protocols under study.

Figure 6 shows some delay spikes regarding SMAC protocol which also is the cause of unsynchronized schedules of the neighboring nodes.

**Packets delivery percentage:** Considering Table 2. We can observe that IEEE 802.11, IEEE 802.15.4. SMAC and TDMA had packet delivery percentage of 97.08%, 92.72%, 30.61% and 93.80%.

As shown in Table 1, traffic of the simulation is set to be generated every 0.1 seconds using constant bit rate traffic generator. Considering the fact that the traffic generation is started at the 10th second, this gives us a total number of 5900 packets which could be transmitted. This indicates that IEEE 802.11, IEEE 802.15.4, SMAC and TDMA transmitted 51.74%, 55.19%, 0.83% and 5.74% of the total possible packets respectively. Considering the packets delivery percentage and the information above, IEEE 802.15.4 and 802.11 outperformed the other two significantly in these criteria.

Table 2. Packet and delay info in 802.11, SMAC, TDMA and 802.15.4 MAC protocols

| | | |
|---|---|---|
| **802.11** | Received Packet Count | 2964 |
| | Dropped Packet Count | 89 |
| | Minimum Delay | 0.009561 |
| | Maximum Delay | 0.529580 |
| | Average Delay | 0.030073 |
| **SMAC** | Received Packet Count | 15 |
| | Dropped Packet Count | 34 |
| | Minimum Delay | 0.271002 |
| | Maximum Delay | 8.084473 |
| | Average Delay | 2.808311 |
| **TDMA** | Received Packet Count | 318 |
| | Dropped Packet Count | 21 |
| | Minimum Delay | 0.460000 |
| | Maximum Delay | 3.058000 |
| | Average Delay | 1.479333 |
| **802.15.4** | Received Packet Count | 3019 |
| | Dropped Packet Count | 237 |
| | Minimum Delay | 0.002482 |
| | Maximum Delay | 0.665996 |
| | Average Delay | 0.019378 |

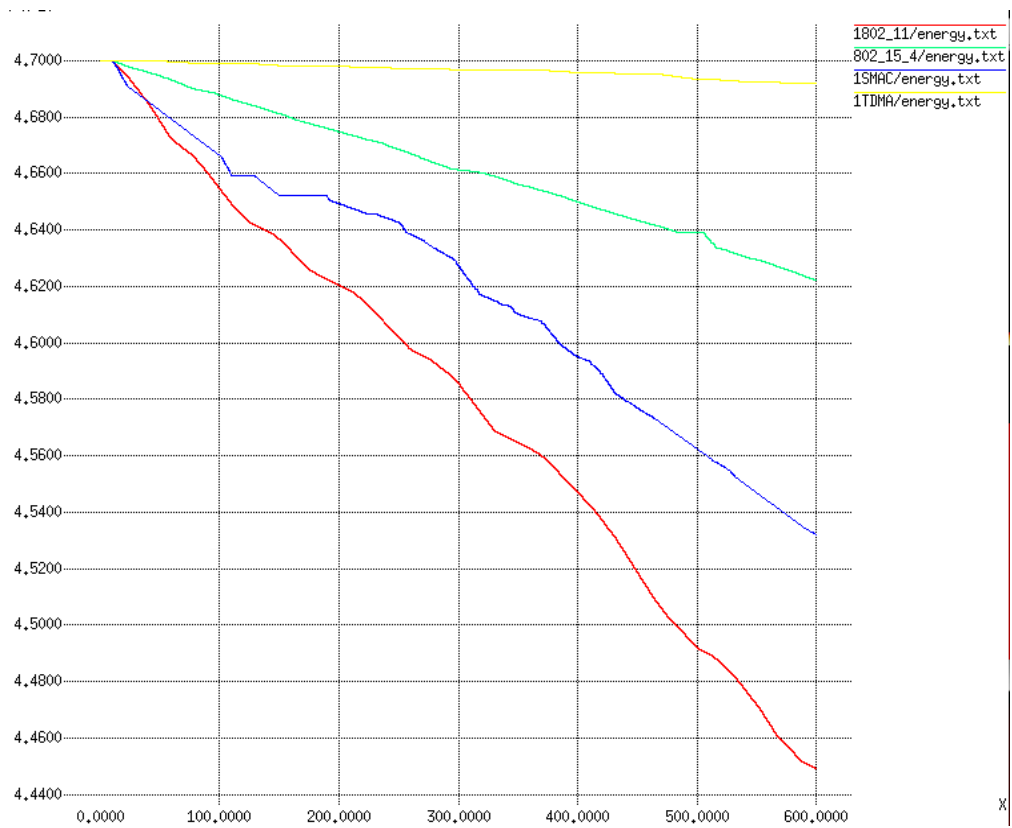

**Figure 5:** Average remaining energy of vehicle node after 600 sec simulations using MAC protocols of 802.15.4, 802.11, SMAC and TDMA

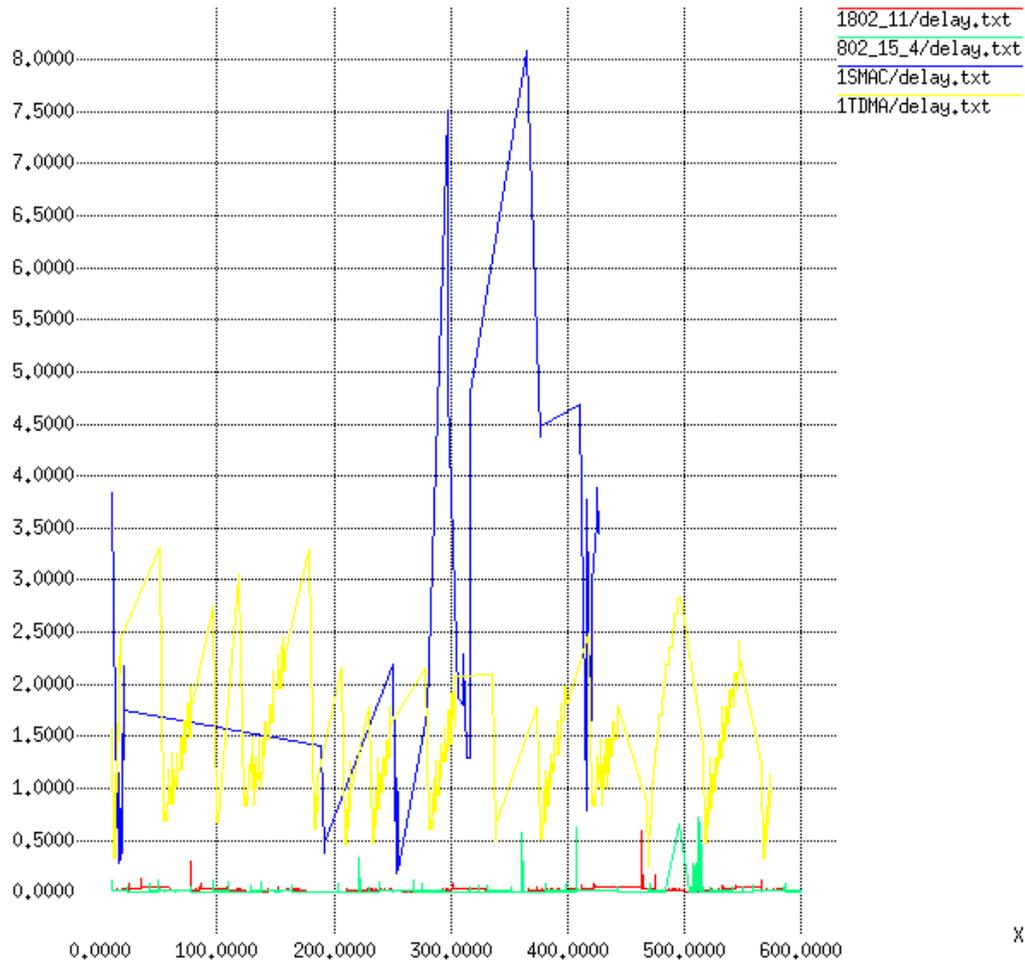

**Figure 6:** Average delay of vehicle node after 600 sec simulations using MAC protocols of 802.15.4, 802.11, SMAC and TDMA

## 7. Conclusions

Lack of automation and enough information in today's environmental pollution regulations is a major obstacle to control the air pollution in large urban areas. In this paper, we have proposed and evaluated a system which uses the help of advances in computer science to overcome such challenges. It is clear that, if we have the required data and leverage to prevent the use of vehicles, which emit pollution more than a standard level, we can drastically reduce the negative impact of such pollution sources. Obviously, in the implementation of gas sensor nodes in modern vehicles, the interval between each data transmission can be much higher than 0.1 second to avoid redundant information to be transmitted more than once in a day. Also, not being able to monitor a small number of vehicles for a while would not have much effect on the whole idea of monitoring gas emission in a large city.

In this paper we compared the performance of four MAC layer protocols of IEEE 802.11, IEEE 802.15.4, SMAC and TDMA. Our results indicated that TDMA had the least energy consumption while total sent packets were much lower than 802.11 or 802.15.4. In aspects of packet delivery, total number of sent packets and average delay, 802.15.4 and 802.11 outperformed the other two. Considering the lower energy consumption of 802.15.4 than 802.11, we choose this protocol to be the best suitable for our exhaust emission monitoring purpose in large cities. Energy consumption of the vehicle node was low enough in all cases for our vehicle node to work for more than a year with power source of 4700 J which is typical power of a normal AA battery which is really suitable for implementation purposes.